# Estimating mesoscale linkage between land-surface conditions and marine productions in Japan: a study using a sparse high-dimensional model

Takeshi Ise[1,2*]
Yurika Oba[1,3]

[1]Field Science Education and Research Center, Kyoto University, Japan
[2]PRESTO, Japan Science and Technology Agency, Japan
[3]Center for the Promotion of Interdisciplinary Education and Research, Kyoto University, Japan
[*]corresponding author: ise@kais.kyoto-u.ac.jp

## Abstract

There have been several scientific studies concerning the interconnectedness between land-surface conditions (e.g., vegetation, land use, and socioeconomic activities) and marine ecosystems (e.g., biodiversity, primary production, and seafood production). This idea of the land-sea connectivity sounded reasonable for many scientists because there is an obvious connection by rivers, however, quantitative estimation of this relationship has been thought to be difficult due to the size of the target areas, the numbers of possible variables, and the amount of noises in this complex system. In this study, we applied a sparse high-dimensional modeling to overcome these difficulties and found several significant land-sea linkages. In this modeling, the key is the penalization in the number of independent variables, and thus a limited number of significant independent variables are chosen in an objective and quantitative manner. We selected 448 independent variables (geological, biological, and social) and 68 dependent variables (marine products) from a governmental database. Then we summarized the data according to the political boundaries of 47 prefectures. The sparse high-dimensional model we constructed successfully highlighted several significant variables to estimate the amount of marine products by prefecture. For example, we found that coastal, especially sessile marine products such as seaweed and shellfish had more explanatory variables than open-water marine products such as tuna and sailfishes. In addition, salmon production had a strong connection to the mesoscale (prefecture-level) land-surface conditions possibly due to their interconnected life cycle between freshwater rivers and the sea. We believe that the sparse modeling is an effective statistical tool to explain relationships in a complex system such as land-sea connections.

# Introduction

There have been several scientific studies concerning the interconnectedness between land-surface conditions (e.g., vegetation, soil type, topology, geochemistry, climate, land use, and socioeconomic activities) and marine ecosystems (e.g., biodiversity, primary production, water quality, and seafood production) in relatively large spatial scales (Gorman et al. 2009, Wang et al. 2010, Waterhouse et al. 2016). This concept of land-sea linkage has attracted ecological and/or economic interests. In Japan, some owners of the aquaculture industry of oysters noticed that the conditions of the forests in the vicinity caused significant impacts on the quality and quantity of the oysters. Thus they made a slogan that says "the sea is longing for the forest (NPO Mori-umi office 2018)" to promote the environmental protection of forest ecosystems for the sake of seafood productions.

This idea of the land-sea connectivity sounded reasonable for many scientists because there is an obvious connection by rivers. However, quantitative estimation of this relationship has been thought to be difficult due to the size of the target areas, the numbers of possible variables, and the amount of noises in this complex natural-anthropogenic system. On land, there are numerous possible variables that have significant effects on marine ecosystems. For example, forest types, agricultural activities, and waste water management may have significant impacts on productions of filter feeders such as oysters. However, the sample size of the study of land-sea linkage is usually limited because observation of large-scale environmental variables is labor-intensive and/or economically expensive. In this case, the problem is often called "high dimension, low sample size" (HDLSS) conditions. In HDLSS conditions, conventional statistical analyses such as regression analysis and analysis of variance (ANOVA) are not useful because of the imbalance between the number of independent variables and the sample size.

Recently, however, a few statistical methods called attention because of their suitability in HDLSS conditions, thanks to the advancements in theory and computational applications in statistics (Fan et al. 2011, Shen et al. 2013). High dimensional sparse modeling is one of these statistical advancements. This approach has mainly been utilized in economics and finance. Using penalty functions, the high dimensional sparse modeling effectively reduces the number of independent variables, and the resultant model is much simpler than the conventional model selection methods such as stepwise AIC.

In this study, we applied a sparse high-dimensional modeling to overcome the difficulties in the study of land-sea linkages and aimed to find significant land variables that affect marine productions. In this modeling, the key is the penalization in the number of independent variables, and thus a limited number of significant independent variables are chosen in an objective and quantitative manner.

# Methods

Our target area is the entire extent of Japan. We selected 448 independent variables (geological, biological, and socioeconomic; Supplementary Table 1) and 68 dependent variables (marine product species or groups; Table 1) from the governmental database (Statistical Observations of Prefectures 2017, Prefectural statistics of fisheries 2016). Then we summarized the data according to the political boundaries of 47 prefectures. We assigned local factors (1-5) to each of the dependent variables (marine product species or groups) according to the degree of sedentary behavior of the species or groups.

The least absolute shrinkage and selection operator (LASSO) is a major statistical technique for the sparse high dimensional modeling. In this study, using a LASSO library glmnet (Friedman et al. 2018) implemented on statistics software R 3.4.4 (R Core Team 2014), we tested statistical relationship between marine products and the mesoscale (prefecture-level) environmental and socioeconomic conditions. LASSO effectively reduced the number of explanatory variables to explain the variance in prefecture-level marine production. In the list of marine products, we took one marine product for each LASSO analysis with all (448) explanatory variables. The degree of penalization in LASSO was optimized by cross validation.

# Results

The sparse high-dimensional model we constructed successfully highlighted several significant variables to estimate the amount of marine products by prefecture. LASSO effectively reduced the number of explanatory variables. With no penalties, all of 448 candidates of the explanatory variables were used to model production of a marine product. Using cross validation, the model systematically changes the degree of penalization to find the optimum. In the optimal condition, most of the 448 candidate variables have a coefficient of zero, which means those variables are not needed in the proposed model. Typically, a few variables have non-zero coefficients (Supplementary Table 1), and that means those variables are statistically significant. In some cases (especially for offshore species), all coefficients are optimized to zero. This indicates that land conditions do not have statistical significance to predict production of that species or groups.

Moreover, our results show ecologically interesting aspects. For example, we found that coastal, especially sessile marine products such as seaweed and shellfish had more explanatory variables than open-water marine products such as tuna and sailfishes. In addition, salmon production had a strong connection to the mesoscale (prefecture-level) land-surface conditions possibly due to their interconnected life cycle between freshwater rivers and the sea.

Table 1. List of dependent variables, assigned local factors, and numbers of explanatory variables.

| Name | local factor | number of explanatory variables |
|---|---|---|
| bluefin tuna | 1 | 0 |
| southern bluefin tuna | 1 | 0 |
| Albacore | 1 | 2 |
| bigeye tuna | 1 | 0 |
| yellowfin tuna | 1 | 0 |
| other tuna | 2 | 6 |
| striped marlin | 1 | 1 |
| swordfish | 1 | 0 |
| black marlin | 1 | 4 |
| other marlins | 2 | 0 |
| bonito | 1 | 0 |
| auxis | 1 | 1 |
| sharks | 2 | 0 |
| salmon | 3 | 22 |
| trout | 3 | 22 |
| gizzard shad | 4 | 0 |
| herring | 3 | 21 |
| sardine | 3 | 5 |
| round herring | 3 | 3 |
| anchovy | 3 | 2 |
| young sardine | 4 | 5 |
| Japanese horse mackerel | 3 | 2 |
| mackerel shad | 2 | 0 |
| mackerels | 2 | 4 |
| Pacific saury | 3 | 15 |
| yellowtail | 2 | 5 |
| flounder | 4 | 1 |
| righteye flounder | 4 | 24 |
| Pacific cod | 4 | 32 |
| walleye pollack | 4 | 15 |
| atka mackerel | 4 | 17 |
| kichiji rockfish | 4 | 1 |
| Japanese sandfish | 3 | 0 |
| deepsea smelt | 4 | 15 |
| conger-eel | 4 | 1 |
| Hairtail | 4 | 33 |
| red sea bream | 3 | 0 |
| crimson sea bream and yellowback sea bream | 3 | 1 |
| black porgy and silver sea bream | 3 | 4 |
| chicken grunt | 4 | 0 |
| Spanish mackerel | 4 | 37 |
| sea bass | 3 | 0 |
| Japanese sand lance | 4 | 0 |
| tilefish | 3 | 0 |
| puffer fish | 4 | 2 |
| other fish species | 3 | 19 |
| lobster | 4 | 0 |
| scampi | 4 | 1 |
| other shrimps | 4 | 0 |
| snow crab | 2 | 0 |
| red snow crab | 2 | 3 |
| swimming crab | 4 | 0 |
| other crabs | 3 | 28 |
| krill | 4 | 30 |
| abalone | 5 | 8 |
| horned turban | 5 | 0 |
| clam | 5 | 34 |
| scallop | 5 | 21 |
| other shellfish | 5 | 37 |
| Japanese common squid | 2 | 0 |
| neon flying squid | 2 | 0 |
| other squid species | 2 | 0 |
| octopus | 4 | 23 |
| sea urchins | 5 | 8 |
| marine mammals | 3 | 3 |
| other aquatic animals | 3 | 1 |
| kelp | 5 | 22 |
| other seaweed | 5 | 0 |

Supplementary Table 1 448 variables of environmental and socioeconomic factors used in the sparse high dimensional analysis. Numbers of times selected as the significant explanatory variable are also indicated.

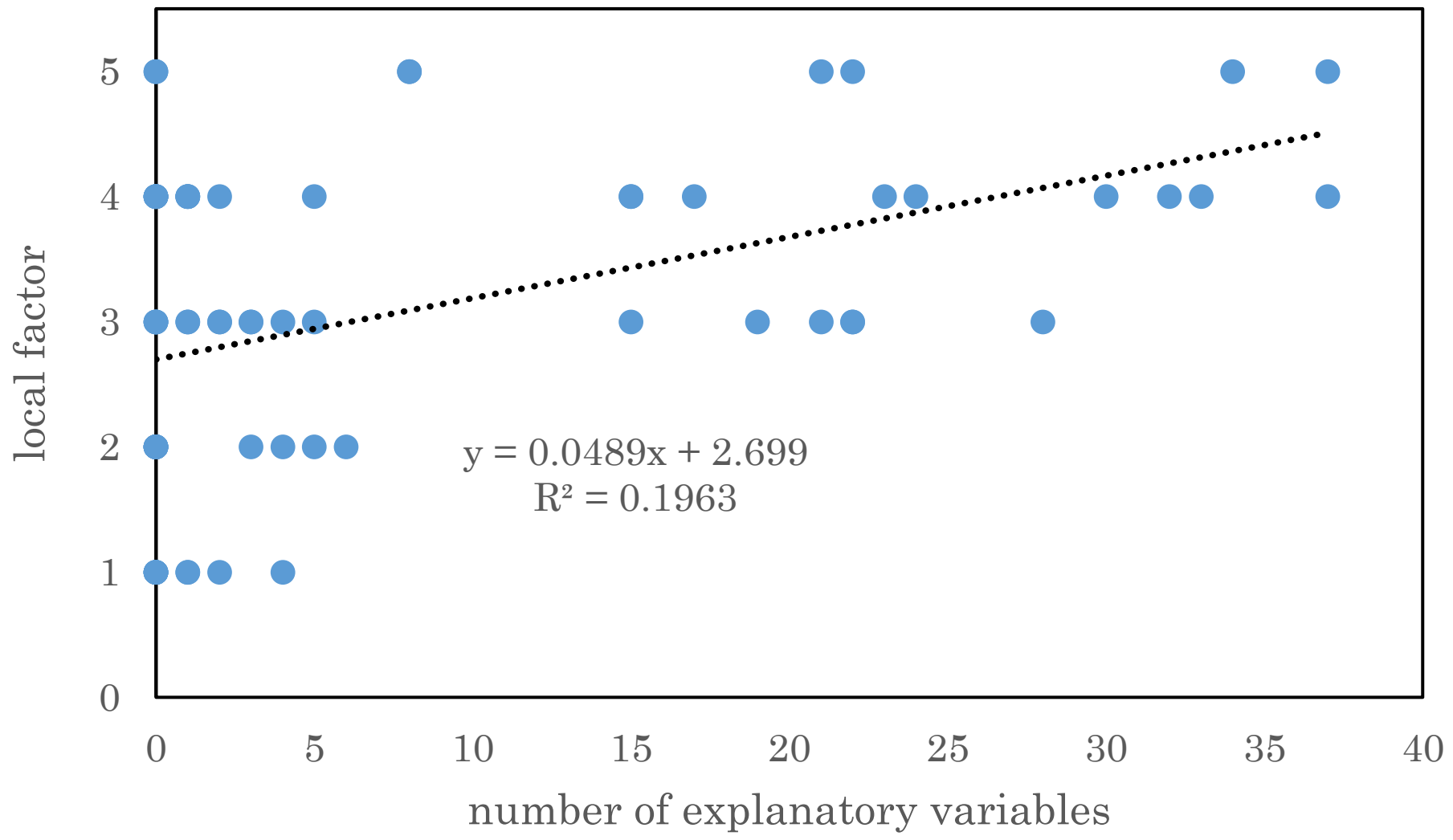

Figure 1. The number of explanatory variables has a significant relationship with the local factor, the degree of connection to the local environments.

# Discussion

In this study, we show that LASSO, a method for sparse high-dimensional modeling, effectively found significant explanatory variables in the mesoscale land-sea linkages. Despite the high dimensionality (448 candidates for the explanatory variables), LASSO worked in a robust manner. We believe that the sparse modeling is an effective statistical tool to explain relationships in a complex system such as land-sea connections.

We found that there are many interesting relationships between land-surface conditions and marine productions in adjacent water. Japan has a relatively large north to south gradient that is reflected in both land-surface conditions and marine productions. For example, Hokkaido, the northernmost prefecture, has a unique characteristics in both land-surface conditions and marine productions. This is obvious in our analysis. For instance, walleye pollack and Atka mackerel, examples of major marine products in Hokkaido, have unique explanatory variables such as household size, the size of agricultural farms, and price indices. We assigned local factors that indicate the degree of ecological connection of the marine products to the local conditions. For example, large, open-water, migratory species such as tuna have low (e.g., 1-2) local factors. On the other hand, small, coastal, sessile species that have limited ability and tendency to move have high (e.g., 4-5) local factors. Species of invertebrates and seaweeds typically have high local factors. A simple regression showed that there is an interesting relationship between the numbers of explanatory variables and the local factors. Species with high local factors tend to have more explanatory variables than that with low local factors. This is intuitive because species with limited ability to move are affected by local conditions, and the local conditions of the sea would be affected by land conditions which is connected by rivers. This finding may show that our LASSO analysis was reasonable.

There are several limitations in the current study. For example, the local factor we assigned for each marine products are somewhat arbitrary. This study is a pure statistical analysis and the mechanistic relationships between land use patterns and marine productions are not revealed in detail. In addition, although we applied LASSO for this study, there are other statistical and computational methods for HDLSS datasets. We are planning to apply the elastic net for the same dataset to see similarities and differences in results. In total, our results encourages that the further studies concerning land-sea interactions are promising and needed.

# Acknowledgments

This study was supported by the Link Again Program of the Nippon Foundation-Kyoto University Joint Project.

Supplementary Table 1

| no. | name | number of dependent variables explained |
|---|---|---|
| 1 | Total population | 0 |
| 2 | Population (male) | 0 |
| 3 | Population (female) | 0 |
| 4 | Foreigner population (per 100 thousand people) | 0 |
| 5 | Population proportion (to the total nationwide population) | 0 |
| 6 | Population density (per 1km2) | 0 |
| 7 | Population density (per 1km2 of inhabitant area) | 0 |
| 8 | Daytime to nighttime population ratio | 0 |
| 9 | Population concentration district population ratio (vs. total population) | 0 |
| 10 | Youth population proportion [under 15 years old] (vs. total population) | 1 |
| 11 | Elderly population proportion [over 65 years old] (vs. total population) | 0 |
| 12 | Production age Population proportion [15 to 64 years old] (vs. total population) | 0 |
| 13 | Youth Population Index (Youth Population / Production Age Population × 100) | 0 |
| 14 | Elderly Population Index (Elderly population / Production age population × 100) | 0 |
| 15 | Dependent population index ((Young + old age population) / Production age population × 100) | 0 |
| 16 | Population change rate ((total population – total population of the previous year) / total population of the previous year) | 0 |
| 17 | Natural rate of change ((Number of births – Deaths) / Total population) | 0 |
| 18 | Rough fertility rate (per thousand people) | 0 |
| 19 | Total special fertility rate | 0 |
| 20 | Rough mortality rate (per thousand people) | 0 |
| 21 | Age-adjusted death rate [man] (per thousand people) | 0 |
| 22 | Age-adjusted death rate [female] (per thousand people) | 0 |
| 23 | Mortality rate by age [0 to 4 years old] (per thousand people) | 0 |
| 24 | Mortality rate by age [over 65 years old] (per thousand population) | 0 |
| 25 | Social change rate ((Number of entrants – Number of out-migrants) / Total population) | 0 |
| 26 | Transfer rate (number of inward migrants / total population) | 0 |
| 27 | Transfer rate (number of out-migrants / total population) | 1 |
| 28 | Inflow population ratio (vs. total population) | 0 |
| 29 | Outflow population ratio (vs. total population) | 1 |
| 30 | Number of general households | 0 |
| 31 | Percentage of general households in nationwide average households | 0 |
| 32 | Average number of people in regular households | 6 |
| 33 | Percentage of nuclear family households (versus general households) | 0 |
| 34 | Percentage of single households (versus general households) | 0 |
| 35 | Percentage of households with household members over 65 years old (number of households in general) | 0 |
| 36 | Percentage of households of only elderly couple (number of general households) | 0 |
| 37 | Percentage of elderly single households (versus general households) | 0 |
| 38 | Dual-income households ratio (versus general households) | 0 |
| 39 | Marriage rate (per thousand people) | 0 |
| 40 | Marriage rate (per thousand people) | 0 |
| 41 | Total area | 11 |
| 42 | Area proportion (total area of ??the whole country) | 2 |
| 43 | Forest area ratio (vs. total area) | 1 |
| 44 | Natural park area ratio (vs. total area) | 1 |
| 45 | Percentage of residential area (vs. total area) | 0 |
| 46 | Annual average temperature | 0 |
| 47 | Maximum temperature (highest monthly average of day maximum temperature) | 3 |
| 48 | The lowest temperature (minimum monthly average of day lowest temperature) | 2 |
| 49 | Year average relative humidity | 1 |
| 50 | Daylight hours (year) | 7 |
| 51 | Rainfall (yearly) | 2 |
| 52 | Sunny days (year) | 2 |
| 53 | Number of rainy days (year) | 0 |
| 54 | Snow days (year) | 0 |
| 55 | Prefecture income per capita | 0 |
| 56 | Gross production in prefecture versus previous year growth rate | 3 |
| 57 | Prefectural gross income [real] vs. previous year growth rate | 0 |
| 58 | Prefectural capital gain versus previous year growth rate | 0 |
| 59 | Prefectural gross income [nominal] vs. previous year growth rate | 1 |
| 60 | Number of secondary industrial establishments (percentage of establishments) | 2 |
| 61 | Number of tertiary industrial establishments (percentage of establishments) | 2 |
| 62 | Percentage of business establishments of 1 to 4 workers (vs. number of private enterprises) | 3 |
| 63 | Percentage of establishments of 100 to 299 employees (vs. number of private enterprises) | 0 |
| 64 | Percentage of establishments with more than 300 employees (vs. number of private enterprises) | 0 |
| 65 | Number of secondary industry workers (per business establishment) | 1 |
| 66 | Number of tertiary industry workers (per business establishment) | 0 |
| 67 | Percentage of employees of business establishments of 1 to 4 employees (number of employees of privately operated businesses) | 0 |
| 68 | Percentage of employees at business establishments of 100 to 299 employees (number of employees at private business establishments) | 2 |
| 69 | Percentage of employees at establishments with 300 or more employees (number of employees at private business establishments) | 0 |
| 70 | Agricultural output per employee [Sales farmer] | 2 |
| 71 | Arable land area ratio (arable land area / total area) | 11 |
| 72 | Land productivity (per hectare of arable land area) | 1 |
| 73 | Arable land area (per farmer) | 18 |
| 74 | Shipment value of manufactured goods etc. (per employee) | 3 |
| 75 | Amount shipped of manufactured goods etc. (per business establishment) | 0 |
| 76 | Commercial annual commodity sales [Wholesale + Retail] (per employee) | 0 |
| 77 | Commercial annual commodity sales [Wholesale business + retail business] (per business establishment) | 0 |
| 78 | Balance of domestic bank deposits (per capita) | 1 |
| 79 | Postal savings outstanding (per capita population) | 6 |
| 80 | National price regional difference index nationwide [comprehensive] | 0 |
| 81 | National price regional difference index [comprehensive] (excluding rent) | 0 |
| 82 | National price regional difference index [Food] | 0 |
| 83 | National price regional difference index [residential] | 0 |
| 84 | National price regional difference index nationwide [Photothermal ? water service] | 2 |
| 85 | National Price Regional Difference Index [Furniture ? Household Goods] | 1 |
| 86 | National price regional difference index [clothing and footwear] | 0 |
| 87 | National price range regional difference index [Health care] | 17 |
| 88 | National price regional difference index nationwide [Traffic & communication] | 1 |
| 89 | National price regional difference index [Education] | 1 |
| 90 | National price regional difference index [cultural entertainment] | 0 |
| 91 | Consumer price regional difference index [Overall: 51 city average = 100] | 0 |
| 92 | Consumer price regional difference index [food: 51 city average = 100] | 4 |
| 93 | Standard price vs. average fluctuation rate over the previous year [residential area] | 1 |
| 94 | Financial strength index [prefectural finance] | 0 |
| 95 | Real income ratio [prefectural finance] | 5 |
| 96 | Percentage of municipal bond current present value (total expenditure settlement accounts) [prefectural finance] | 0 |
| 97 | Current account ratio [prefectural finance] | 5 |
| 98 | Percentage of voluntary financial resources (total expenditure settlement accounts) [prefectural finance] | 0 |
| 99 | Percentage of general resources (total expenditure settlement accounts) [prefectural finance] | 1 |
| 100 | Ratio of investment expenses (total expenditure settlement accounts) [prefectural finance] | 1 |
| 101 | Local tax rate (total revenue settlement accounts) [prefectural finance] | 0 |
| 102 | Local allocation tax rate (total revenue settlement accounts) [prefectural finance] | 0 |
| 103 | Treasury expenditure ratio (total revenue settlement accounts) [prefectural finance] | 1 |
| 104 | Residential tax (per capita population) [prefecture / municipal finance total] | 0 |
| 105 | Property tax (per capita population) [Prefecture / Municipal Finance Total] | 0 |
| 106 | Taxable income (per taxpayer) | 0 |
| 107 | Percentage of consumer expenditure (total expenditure settlement accounts) [prefectural finance] | 2 |
| 108 | Social welfare expenditure ratio (total expenditure settlement accounts) [prefectural finance] | 0 |
| 109 | Old welfare expenditure ratio (total expenditure settlement accounts) [prefectural finance] | 1 |
| 110 | Child welfare expenditure ratio (total expenditure settlement accounts) [prefectural finance] | 3 |
| 111 | Life protection expenditure ratio (total expenditure settlement accounts) [prefectural finance] | 0 |
| 112 | Percentage of sanitation expenses (total expenditure settlement accounts) [prefectural finance] | 0 |
| 113 | Percentage of labor costs (total expenditure settlement accounts) [prefectural finance] | 1 |
| 114 | Ratio of agriculture, forestry and fishery industry expenditure (total expenditure settlement accounts) [prefectural finance] | 0 |
| 115 | Commerce and construction expenditure ratio (total expenditure settlement accounts) [prefectural finance] | 0 |
| 116 | Civil engineering expenditure ratio (total expenditure settlement accounts) [prefectural finance] | 2 |
| 117 | Percentage of police expenses (total expenditure settlement accounts) [prefectural finance] | 0 |
| 118 | Fire fighting proportion (total expenditure settlement accounts) [municipal finance] <including city> | 2 |
| 119 | Percentage of education expenditure (total expenditure settlement accounts) [prefectural finance] | 0 |
| 120 | Disaster recovery expenditure ratio (total expenditure settlement accounts) [prefectural finance] | 3 |
| 121 | Personnel expenditure ratio (total expenditure settlement accounts) [prefectural finance] | 0 |
| 122 | Relief expenditure ratio (total expenditure settlement accounts) [prefectural finance] | 2 |
| 123 | Normal Construction Project Cost Ratio (Total Expenditure Results) [Prefectural Finance] | 0 |
| 124 | Total annual expenditure account (per capita) [prefecture ? municipal finance total] | 0 |
| 125 | Private Expenditure (per capita population) [Prefecture / Municipal Finance Total] | 4 |
| 126 | Social welfare expenditure (per capita population) [prefecture / municipal fiscal total] | 1 |
| 127 | Elderly welfare expenses (Per capita over 65 years old) [Prefecture ? municipal fiscal total] | 0 |
| 128 | Child welfare expenses (per capita population aged 17 or under) [Prefectural / Municipal fiscal total] | 0 |
| 129 | Living conservation expenses (Per capita for protected real persons) [Prefecture / Municipal Finance Total] | 0 |
| 130 | Sanitation expenditure (per capita population) [prefecture / municipal finance total] | 0 |
| 131 | Civil engineering expenditure (per capita population) [prefecture / municipal finance total] | 0 |
| 132 | Police charges (per capita population) [prefectural finance] | 0 |
| 133 | Firefighting expenses (per capita) [Total for municipalities / municipalities] | 1 |
| 134 | Educational expenditure (per capita population) [prefecture / municipal finance total] | 0 |
| 135 | Social education fee (per capita) [prefecture / municipal finance total] | 1 |
| 136 | Disaster restoration cost (per capita) [Total for municipalities / municipalities] | 6 |
| 137 | Public elementary school expenses (per child) [Prefecture ? municipal finance total] | 0 |
| 138 | Public junior high school expenses (per student) [prefecture / municipal fiscal total] | 0 |
| 139 | Public high school expenses (per student) [Prefectural / municipal fiscal total] | 1 |
| 140 | Special school fee <public> (per student / student) [prefecture / municipal finance total] | 1 |
| 141 | Kindergarten fee (per child) [Prefectural / Municipal Finance Total] | 0 |
| 142 | Number of elementary schools (per 6 million to 11 years of population per 100,000 people) | 0 |
| 143 | Number of junior high school (per 100,000 people ages 12 to 14) | 0 |
| 144 | Number of high schools (per 100,000 people aged 15 to 17) | 0 |
| 145 | Number of kindergartens (per 3 0 to 5 years old population of 100,000) | 7 |
| 146 | Number of nursery schools (per 0 0 5 population per 100 000 people) | 0 |
| 147 | Number of elementary schools (per 100 kilometers of permanent area) | 0 |
| 148 | Middle school (per 100 kilometers of permanent area) | 0 |
| 149 | Number of high schools (per 100 kilometers of permanent area) | 0 |
| 150 | Elementary school girls teacher ratio (vs. elementary school teachers) | 4 |
| 151 | Junior high school girls teacher ratio (number of middle school teachers) | 1 |
| 152 | Number of elementary school children (per elementary school teacher) | 0 |
| 153 | Number of middle school students (per teacher in junior high school) | 0 |
| 154 | Number of high school students (per high school faculty member) | 0 |
| 155 | Number of kindergarten residents (per kindergarten teacher) | 0 |
| 156 | Number of childcare centers (per childcare day nurser) | 2 |
| 157 | Public high school student ratio (number of high school students) | 0 |
| 158 | Ratio of public kindergarten visitors (against the number of kindergarten visitors) | 6 |
| 159 | Percentage of Public Daycare Children's Children (vs. Day Nurseries) | 4 |
| 160 | Number of elementary school children (per class) | 0 |
| 161 | Number of middle school students (per class) | 0 |
| 162 | Kindergarten education penetration level (Number of graduates who completed kindergartens / Number of elementary school children | 0 |
| 163 | Nursery school education popularity degree (number of people who completed nursery school / number of elementary school children | 1 |
| 164 | Elementary school long absentee child absence rate due to school refusal (per thousand children) 1) | 1 |
| 165 | Middle school long absentee student ratio by school refusal (per thousand students) 1) | 3 |
| 166 | The rate of advancement of junior high school graduates | 6 |
| 167 | The rate of advancement of senior high school graduates | 0 |
| 168 | Number of universities (per 100 thousand people) | 3 |
| 169 | Ratio of entrants to universities in the high school where the high school is located (number of university entrants to university) | 2 |
| 170 | College capacity index (number of high school graduates entering university) | 0 |
| 171 | Junior college number (per 100 thousand people) | 4 |
| 172 | Number of vocational schools (per 100 thousand people) | 1 |
| 173 | Number of schools (per 100 thousand people) | 7 |
| 174 | Percentage of those who graduated from elementary and junior high school with the final academic background (total number of graduates) | 0 |
| 175 | Percentage of high school and former middle school graduates with final academic background (total number of graduates) | 0 |
| 176 | Percentage of those who graduated from junior college / technical college degree (total number of graduates) | 0 |
| 177 | Percentage of graduate / graduate graduate graduate graduate graduate graduate graduate graduate graduate graduate final academic history | 0 |
| 178 | Elementary school education fee (per child) | 0 |
| 179 | Elementary school education fee (per student) | 0 |
| 180 | High school education fee [Full time] (per student) | 0 |
| 181 | Kindergarten education expenditure (per visitor) | 0 |
| 182 | Labor force population ratio (vs. 15 years old population) [male] | 1 |
| 183 | Labor force population ratio (vs. 15 years old population) [female] | 0 |
| 184 | Percentage of primary industry workers (vs. workers) | 0 |
| 185 | Secondary industrial workers ratio (vs employed) | 0 |
| 186 | Third Industrial Employee Ratio (Versus Employee) | 2 |
| 187 | Complete unemployment rate (total number of unemployed people / labor force population) | 0 |
| 188 | Employment ratio (Number of employees / Employees) | 2 |
| 189 | Percentage of employed workers in the prefecture (vs. workers) | 0 |
| 190 | Percentage of commuters to other municipalities, towns and villages (vs. workers) | 0 |
| 191 | Percentage of commuters from other municipalities, towns and villages (vs. workers) | 0 |
| 192 | Employment rate (number of employment / number of job seekers) | 0 |
| 193 | Effective job openings ratio (number of job openings / number of job seekers) | 2 |
| 194 | Sufficiency rate (number of jobs / number of job openings) | 5 |
| 195 | Part time Employment rate [regular] (Number of employment / Number of job seekers) | 2 |
| 196 | Middle-aged and elderly employment Employment rate [over 45 years old] (number of employment / number of candidates) | 0 |
| 197 | Middle-aged and elderly employment ratio [over 45 years old] (number of employment versus employment) | 2 |
| 198 | Percentage of older workers [over 65 years old] (Old age population) | 3 |
| 199 | Percentage of older general workers [over 65 years old] (Old age population) | 4 |
| 200 | Percentage of people with disabled employment (per thousand of employment opportunities) | 2 |
| 201 | Percentage of employment in high school graduates (number of high school graduates) | 0 |
| 202 | Percentage of employees outside the prefecture occupying high school graduates (number of high school graduates employed) | 6 |
| 203 | Margin ratio of new high school graduates (versus new high school graduate job seekers) | 0 |
| 204 | Percentage of employment in university graduates (number of university graduates) | 3 |
| 205 | Unemployed rate of university new graduates (number of university graduates) | 1 |
| 206 | Rate of job change (number of job change / number of contractors) | 0 |
| 207 | Retirement rate (number of departures / (number of continuing workers + number of change workers + number of departures)) | 0 |
| 208 | New employment rate (number of newly employed workers / number of contractors) | 0 |
| 209 | Employment transfer rate ((Number of job change + number of people leaving + number of newly employed people) / Population aged 15 years or over) | 0 |
| 210 | Actual working hours (Month) [Man] | 4 |
| 211 | Actual working hours (Month) [F] | 1 |
| 212 | Male part-time salary (per hour) | 0 |
| 213 | Female part-time salary (per hour) | 0 |
| 214 | Male part-time workers | 0 |
| 215 | Number of female part-time workers | 0 |
| 216 | New high school newly graduated starting salary (monthly) [male] | 0 |
| 217 | New high school newly graduated starting salary (monthly) [woman] | 0 |
| 218 | Public private hall number (per 1 million people) | 0 |
| 219 | Number of libraries (per 1 million people) | 0 |
| 220 | Number of museums (per 1 million people) | 0 |
| 221 | Number of youth education facilities (per 1 million people) | 2 |
| 222 | Permanent movie theater (per million population) | 3 |
| 223 | Number of social sports facilities (per 1 million people) | 0 |
| 224 | Multipurpose exercise plaza (per million population) | 1 |
| 225 | Youth classes / number of courses (per 1 million people) | 1 |
| 226 | Adult general class ? Number of courses (per 1 million people) | 4 |
| 227 | Female class ? Number of courses (per 1 million female population) | 2 |
| 228 | Elderly Classes ? Number of Courses (per 1 million population) | 1 |
| 229 | Annual participant rate of volunteer activities (over 10 years old) | 1 |
| 230 | Annual behavior rate of sports (10 years old and over) | 0 |
| 231 | Annual followers rate of travel and holiday (over 10 years old) | 1 |
| 232 | Annual followers rate of overseas travel (over 10 years old) | 0 |
| 233 | Room occupancy rate | 1 |
| 234 | Number of general passports issued (per thousand people) | 0 |
| 235 | Construction New housing ratio (number of houses with resident households) | 0 |
| 236 | House ownership ratio (number of houses with resident households) | 0 |
| 237 | Ratio ratio (number of houses with resident households) | 0 |
| 238 | Private tenancy ratio (number of houses with resident households) | 0 |
| 239 | Ratio of vacancies (vs. total number of houses) | 6 |
| 240 | Construction New Owner-occupancy Ratio (Number of Newly Established Housing Started to Work) | 0 |
| 241 | Construction newly rented house ratio (number of new construction starts to construction) | 0 |
| 242 | Single-family housing ratio (number of houses with resident households) | 0 |
| 243 | Common housing ratio (number of houses with resident households) | 0 |
| 244 | Land area of ??housing (per housing) | 9 |
| 245 | Total area of ??owner-occupied houses (per housing) | 0 |
| 246 | Total area of ??rented houses (per housing) | 0 |
| 247 | Tatami number of owned house (per housing) | 2 |
| 248 | Tatami number of rented housing (per housing) | 1 |
| 249 | Floor area of ??newly-built house started (per housing) | 1 |
| 250 | Floor area of ??newly constructed rental housing (per housing) | 3 |
| 251 | Number of living rooms (per housing) <Owner's house> | 0 |
| 252 | Number of living rooms (per housing) <rented house> | 1 |
| 253 | Tatami number of owned house (per person) | 0 |
| 254 | Tatami number of rented housing (per person) | 5 |
| 255 | Housing ratio with flush toilets (number of houses with resident households) | 1 |
| 256 | House ratio with bathroom (number of houses with resident households) | 0 |
| 257 | Minimum residence area level or more Households proportion | 0 |
| 258 | Percentage of regular households who mainly support households are employers [Commuting time 90 minutes or more] 1) | 0 |
| 259 | Rent for public rental housing (per 3.3 square meters per month) | 0 |
| 260 | Rent of private rental housing (per 3.3 square meters per month) | 1 |
| 261 | Estimated planned construction expenses for buildings (per 1 m2 of floor area) | 1 |
| 262 | Household ratio in city gas supply area (versus general households) | 0 |
| 263 | City gas sales volume | 1 |
| 264 | Gasoline sales volume | 0 |
| 265 | Water supply population ratio of water supply | 4 |
| 266 | Sewerage penetration rate | 4 |
| 267 | Urine treatment population ratio | 0 |
| 268 | Garbage recycling rate | 1 |
| 269 | Landfill disposal rate | 3 |
| 270 | Final disposal site residual capacity | 2 |
| 271 | Number of retail stores (per thousand people) | 0 |
| 272 | Number of large retailers (per 100 thousand people) | 0 |
| 273 | Department stores, total number of supermarkets (per 100 thousand people) | 2 |
| 274 | Number of self-service establishments (per 100 thousand people) | 0 |
| 275 | Number of convenience stores (per 100 thousand people) | 0 |
| 276 | Number of restaurants (per thousand people) | 1 |
| 277 | Number of barbers and beauty shops (per 100 thousand people) | 0 |
| 278 | Number of cleaning stations (per 100 thousand people) | 3 |
| 279 | Number of public baths (per 100 thousand people) | 0 |
| 280 | Number of post offices (per 100 kilometers of permanent area) | 1 |
| 281 | Number of residential telephone subscriptions (per thousand population) | 1 |
| 282 | Mobile phone subscriptions (per thousand people) | 1 |
| 283 | Actual extension of road (per 1 kilometer area) | 0 |
| 284 | Actual extension of main road (per 1 kilometer area) | 1 |
| 285 | Main road pavement ratio (actual extension of major roads) | 1 |
| 286 | Municipal road pavement ratio (vs. municipal road actual extension) | 0 |
| 287 | Number of vehicles owned (per thousand people) | 0 |
| 288 | Number of passenger cars for private use (per thousand people) | 0 |
| 289 | Urbanization adjustment area area ratio (town planning area designated area) | 0 |
| 290 | Residential area area ratio (versus area of ??application area) | 1 |
| 291 | Industrial area exclusive area ratio (versus area of ??application area) | 3 |
| 292 | Urban park area (per capita population) | 3 |
| 293 | Urban parks (per 100 kilometers of permanent area) | 0 |
| 294 | Rate of appeal (per thousand people) | 2 |
| 295 | Visiting rate (per thousand people) | 1 |
| 296 | General hospital Number of new hospitalized patients annually (per 100 thousand people) | 1 |
| 297 | The average number of outpatients per day for general hospitals (per 100 thousand people) | 0 |
| 298 | The average number of in-hospital patients a day at general hospitals (per 100 thousand people) | 0 |
| 299 | Standardized mortality rate (base population = Showa 5 years) (per thousand population) | 0 |
| 300 | Average extra life [0 years old ? man] | 0 |
| 301 | Average extra life [0 years old ? female] | 0 |
| 302 | Average extra life [65 years old ? man] | 1 |
| 303 | Average excess life [65 years old ? female] | 0 |
| 304 | Number of deaths due to lifestyle diseases (per 100 thousand people) | 0 |
| 305 | Deaths due to malignant neoplasms (per 100 thousand people) | 0 |
| 306 | Death due to diabetes (per 100 thousand people) | 1 |
| 307 | Deaths due to hypertensive disease (per 100 thousand people) | 2 |
| 308 | Number of deaths due to heart disease [excluding high blood pressure] (per 100,000 population) | 0 |
| 309 | Death due to cerebrovascular disease (per 100 thousand people) | 1 |
| 310 | Mortality from pregnancy, parturition and production (not including obstetric tetanus) | 3 |
| 311 | Death rate (number of stillbirth / (number of births + stillbirths)) (per thousand births) | 3 |
| 312 | Perinatal mortality rate ((number of stillbirths (after 22 weeks of pregnancy) + number of early neonatal deaths) / | 0 |
| 313 | Neonatal mortality rate (neonatal death / live number) (per thousand of births) | 3 |
| 314 | Infant mortality rate (infant mortality / number of births) (per thousand births) | 0 |
| 315 | Less than 2,500 g Fertility rate (Less than 2,500 g / Number of births) 1) | 3 |
| 316 | Average body length (2nd year junior high school student) | 0 |
| 317 | Average length (2nd year junior high school girl) | 0 |
| 318 | Average weight (2nd year junior high school student) | 1 |
| 319 | Average weight (2nd year junior high school girl) | 9 |
| 320 | Number of general hospitals (per 100 thousand people) | 0 |
| 321 | Number of general clinics (per 100 thousand people) | 1 |
| 322 | Number of psychiatric hospitals (per 100 thousand people) | 0 |
| 323 | Number of dental clinics (per 100 thousand people) | 1 |
| 324 | Number of general hospitals (per 100 kilometers of permanent area) | 0 |
| 325 | Number of general clinics (per 100 kilometers of permanent area) | 0 |
| 326 | Number of dental clinics (per 100 kilometers of permanent area) | 0 |
| 327 | Number of general hospital beds (per 100 thousand people) | 0 |
| 328 | Number of mental beds (per 100 thousand people) | 1 |
| 329 | Number of nursing care medical type medical facilities (per 100,000 people over 65) | 1 |
| 330 | Number of doctors engaged in medical facilities (per 100 thousand people) | 6 |
| 331 | Number of dentists engaged in medical facilities (per 100 thousand people) | 0 |
| 332 | Number of nurses and quasi-nurses engaged in medical facilities (per 100 thousand people) | 0 |
| 333 | General hospital full-time physician number (100 hospital beds) | 0 |
| 334 | Number of general hospital nurses / associate nurses (100 hospital beds) | 0 |
| 335 | Number of general hospital outpatients (full-time doctor per person per day) | 1 |
| 336 | Number of patients in general hospital stay (full-time doctor per person per day) | 0 |
| 337 | Number of patients in general hospital in hospital (nurses / associate nurses per person per day) | 0 |
| 338 | General hospital bed use rate (total number of hospital patients / general number of beds total) | 0 |
| 339 | General hospital average hospital stay (per inpatient patient) | 0 |
| 340 | Number of public health nurses (per 100 thousand people) | 0 |
| 341 | Number of emergency notification hospitals / general clinics (per 100 thousand people) | 3 |
| 342 | Number of ambulances (per 100 thousand people) | 0 |
| 343 | Annual emergency number of cases of emergency (per thousand people) | 1 |
| 344 | Number of drug stations (per 100 thousand people) | 1 |
| 345 | Number of drug stations (per 100 kilometers of permanent area) | 0 |
| 346 | Number of pharmaceutical sales business (per 100 thousand people) | 1 |
| 347 | Number of pharmaceutical sales companies (per 100 kilometers of permanent area) | 0 |
| 348 | Real life protected personnel (per thousand population) | 0 |
| 349 | Life protection education assistance staff (per thousand people) | 9 |
| 350 | Life protection medical assistance staff (per thousand people) | 1 |
| 351 | Life protection Housing assistance staff (per thousand people) | 0 |
| 352 | Life protection nursing care assistant (per thousand people) | 1 |
| 353 | Number of elderly people protected for welfare (per 65 thousand people or more per 1,000 people) | 0 |
| 354 | Number of handbooks handed out for disabled (per thousand people) | 0 |
| 355 | Number of protected facilities (excluding medical protection facilities) (Actual protection workers for welfare protection) | 1 |
| 356 | Number of senior citizens (per person over 65 years old per 100 thousand people) | 0 |
| 357 | Number of elderly welfare centers (per 100,000 people over 65) | 3 |
| 358 | Number of nursing-home aged welfare facilities (per 100,000 people over 65) | 0 |
| 359 | Number of child welfare facilities (per 100 thousand people) | 2 |
| 360 | Number of people living protection facilities Capacity (per thousand of real people protected by welfare) | 1 |
| 361 | Number of residents in welfare facilities (per thousand of real people protected by welfare) | 0 |
| 362 | Elderly Home Capacity (per thousand population over 65) | 3 |
| 363 | Number of nursing home residents (per thousand people over 65) | 0 |
| 364 | Number of committee members (child committee members) (per 100,000 population) | 0 |
| 365 | Maternal and child independence support number (per 100 thousand people) | 2 |
| 366 | Number of visiting nursing care users (per visiting nursing care facility) | 0 |
| 367 | Consumer Commissioner (Children's Committee) Number of counseling and support 1) | 3 |
| 368 | Child consultation center reception number (per thousand people) | 3 |
| 369 | National medical expenditure per capita | 0 |
| 370 | Medical expenses for late-stage elderly people (per insured person) | 0 |
| 371 | Number of National Pension Insured [No. 1] (per thousand of 20 to 59 year old population) | 0 |
| 372 | Number of National Pension Insured [No. 3] (per thousand of 20 to 59 year old population) | 2 |
| 373 | National health insurance number of insured persons (per thousand people) | 0 |
| 374 | National health insurance examination rate (per thousand insured) | 1 |
| 375 | National health insurance medical expense (per insured person) | 0 |
| 376 | Number of health insurance subscribers administered by the National Health Insurance Association (per 1,000 people) | 1 |
| 377 | National Health Insurance Association administered health insurance examination rate (per thousand insured) | 0 |
| 378 | National Health Insurance Association administered health insurance examination rate (per thousand of dependents) | 2 |
| 379 | National Health Insurance Association administered health insurance medical expenses (per insured person) | 2 |
| 380 | National Health Insurance Association administered health insurance medical expenses (per dependent person) | 3 |
| 381 | Employment Insurance Receiving Ratio (Number of Insured People) | 0 |
| 382 | Employee accident compensation insurance benefit rate (vs. number of applicable workers) | 3 |
| 383 | Frequency of occupational accidents | 5 |
| 384 | Degree of weight of occupational accidents | 3 |
| 385 | Number of fire stations (per 100 kilometers of permanent area) | 0 |
| 386 | Fire Department ? Number of branches (per 100 kilometers of permanent area) | 0 |
| 387 | Fire-fighting pumps etc. Current number of vehicles (per 100 thousand people) | 9 |
| 388 | Number of fire brigade (per 100 thousand people) | 2 |
| 389 | Number of firefighters (per 100 thousand people) | 0 |
| 390 | Fire-fighting institution number of times (per 100 thousand people) | 2 |
| 391 | Fire-fighting fire extinguishing count for fire (per 100 thousand people) | 5 |
| 392 | Number of fire fires (per 100 thousand people) | 3 |
| 393 | Number of building fire fires (per 100 thousand people) | 1 |
| 394 | Number of fire casualties (per 100 thousand people) | 0 |
| 395 | Number of fire casualties (per 100 building fires) | 3 |
| 396 | Building Fire Damage Amount (Per Capita) | 2 |
| 397 | Building fire damage amount (per building fire) | 1 |
| 398 | Number of solid crossing facilities (per actual thousand km of road actual extension) | 0 |
| 399 | Number of crosswalks (per actual thousand km) | 0 |
| 400 | Number of traffic lights installed (per 1 thousand km of road actual extension) | 0 |
| 401 | Number of traffic accidents (per thousand km extent of road actual extension) | 0 |
| 402 | Number of traffic accidents (per 100 thousand people) | 1 |
| 403 | Traffic accident casualties (per 100 thousand people) | 0 |
| 404 | Traffic accident deaths (per 100 thousand people) | 2 |
| 405 | Number of cases cleared for road traffic law (per thousand people) | 1 |
| 406 | Police station, police station, number of residential stations (per 100 kilometers of residential area) | 0 |
| 407 | Number of police officers (per thousand people) | 0 |
| 408 | Number of penal code offenses (per thousand people) | 1 |
| 409 | Number of thieves recognized (per thousand people) | 1 |
| 410 | Penal code offense clearance rate (per perceived number) | 0 |
| 411 | Thiever clearance rate (per perceived number) | 1 |
| 412 | Amount of disaster damage (per capita population) | 6 |
| 413 | Number of deaths due to accident (per 100 thousand people) | 1 |
| 414 | Number of pollution complaints (per 100 thousand people) | 2 |
| 415 | Number of smoke generating facilities | 2 |
| 416 | Number of general dust generation facilities | 4 |
| 417 | Specific branch office under the Water Pollution Control Law | 4 |
| 418 | Number of private life insurance holding contracts (per thousand people) | 1 |
| 419 | Private life insurance policy amount (per in-force contract) | 0 |
| 420 | Private life insurance policy amount (per household) | 0 |
| 421 | Fire Insurance Housing Property ? General Property Number of new contracts (per household of 1,000 households) | 1 |
| 422 | Fire insurance housing property ? General property receipt insurance amount (per ownership contract) | 6 |
| 423 | Actual income (one month per household) [Worker households] | 1 |
| 424 | Household head income (1 month per household) [worker households] | 1 |
| 425 | Consumption expenditure (one month per household) [Household of two or more people] | 1 |
| 426 | Food expenditure ratio (vs. consumption expenditure) [Household of two or more people] | 0 |
| 427 | Housing cost ratio (vs. consumption expenditure) [Household of two or more people] | 1 |
| 428 | Light heat / water cost ratio (vs. consumption expenditure) [Household of two or more people] | 4 |
| 429 | Furniture ? Household goods expenses ratio (against consumption expenditure) [Household with two or more people] | 2 |
| 430 | Clothing and footwear cost ratio (vs. consumption expenditure) [Household of two or more people] | 3 |
| 431 | Health care expenditure ratio (vs. consumption expenditure) [Household of two or more people] | 5 |
| 432 | Traffic / communication cost ratio (vs. consumption expenditure) [Household of two or more people] | 1 |
| 433 | Educational expenditure ratio (vs. consumption expenditure) [Household of two or more people] | 2 |
| 434 | Educational entertainment expenses ratio (vs. consumption expenditure) [Household of two or more people] | 0 |
| 435 | Average propensity to consume (consumption expenditure / disposable income) [worker households] | 3 |
| 436 | Current savings (per household) [Household of two or more people] | 6 |
| 437 | Saving deposit Current high percentage (vs. current savings) [Household of two or more people] | 3 |
| 438 | Life insurance Current high ratio (vs. savings present high) [Household of two or more people] | 1 |
| 439 | Current high value of securities (high relative to savings) [Household of two or more people] | 0 |
| 440 | Current liability (per household) [Household of two or more people] | 0 |
| 441 | Debt ratio for housing / land (current debt current) [Household with two or more people] | 0 |
| 442 | Car owned quantity (per thousand households) [Household of two or more people] | 0 |
| 443 | Quantity owned by microwave oven (including electronic oven range) (per thousand households) | 1 |
| 444 | Room air conditioner owned quantity (per thousand households) [two or more households] | 2 |
| 445 | Stereoset or CD ? MD radio cassette owned quantity (per thousand households) | 0 |
| 446 | Number of pianos owned (per 1,000 households) [Household of two or more people] | 1 |
| 447 | Cellular phone (including PHS) owned quantity (per thousand households) [two or more households] | 1 |
| 448 | Number of personal computers owned (per 1,000 households) [Household of two or more people] | 0 |